\documentclass[12pt]{article}

\usepackage{graphicx}
\usepackage{a4}

\usepackage{fancybox}
\usepackage{epsfig}
\usepackage{color}

\usepackage{amsfonts}
\usepackage{mathrsfs}
\usepackage{amssymb}
\usepackage{amsmath}

\usepackage{dcolumn}
\usepackage{bm}

\def\pa{\partial}

\def\al{\alpha}
\def\be{\beta}

\def\de{\delta}

\def\th{\theta}

\def\om{\omega}

\newcommand{\ben}{\begin{equation}}
\newcommand{\een}{\end{equation}}
\newcommand{\bea}{\begin{eqnarray}}
\newcommand{\eea}{\end{eqnarray}}
\newcommand{\ba}{\begin{array}}
\newcommand{\ea}{\end{array}}
\newcommand{\bit}{\begin{itemize}}
\newcommand{\eit}{\end{itemize}}

\newcommand{\vs}[1]{\vspace{#1 mm}}

\newcommand{\dsl}{\pa \kern-0.5em /}

\begin{document}

\topmargin 0pt \oddsidemargin 0mm

\begin{flushright}

USTC-ICTS-11-16 \\

\end{flushright}

\vspace{2mm}

\begin{center}

{\Large \bf Tachyon cosmology with non-vanishing minimum potential:
a unified model}

\vs{10}

 {\large Huiquan Li \footnote{E-mail: hqli@ustc.edu.cn}}

\vspace{6mm}

{\em

Interdisciplinary Center for Theoretical Study\\

University of Science and Technology of China, Hefei, Anhui 230026, China\\

}

\end{center}

\vs{9}

\begin{abstract}
We investigate the tachyon condensation process in the effective
theory with non-vanishing minimum potential and its implications to
cosmology. It is shown that the tachyon condensation on an unstable
three-brane described by this modified tachyon field theory leads to
lower-dimensional branes (defects) forming within a stable
three-brane. Thus, in the cosmological background, we can get
well-behaved tachyon matter after tachyon inflation, (partially)
avoiding difficulties encountered in the original tachyon
cosmological models. This feature also implies that the tachyon
inflated and reheated universe is followed by a Chaplygin gas dark
matter and dark energy universe. Hence, such an unstable three-brane
behaves quite like our universe, reproducing the key features of the
whole evolutionary history of the universe and providing a unified
description of inflaton, dark matter and dark energy in a very
simple single-scalar field model.
\end{abstract}



\section{Introduction}
\label{sec:introduction}

Rolling tachyon on unstable D-branes is thought of a natural
candidate to derive inflation
\cite{Gibbons:2002md,Fairbairn:2002yp,Choudhury:2002xu,Frolov:2002rr,
Kofman:2002rh,Piao:2002vf,Steer:2003yu}. This tachyon inflation
model has some particular merits, e.g., the inflation can end up
with a radiation-dominated stage \cite{Cline:2002it,Barnaby:2004dp}
and a ``tachyon matter"-dominated stage
\cite{Sen:2002in,Sen:2002an,Shiu:2002qe}. All the processes can be
described by the Dirac-Born-Infeld (DBI) effective action of a
rolling tachyon \cite{Garousi:2000tr,Bergshoeff:2000dq,Sen:2002an}
in the cosmological background:
\begin{equation}\label{e:TDBIA}
 S=-\int d^4x\sqrt{-g}V(T)\sqrt{1+g^{\mu\nu}
\partial_\mu{T}\partial_\nu{T}},
\end{equation}
where $V(T)$ is the potential. Besides other forms, the tachyon
potential derived in open string field theory in
\cite{Kutasov:2003er} is
\begin{equation}\label{e:pot}
 V(T)=\frac{V_m}{\cosh(\be T)},
\end{equation}
where $V_m$ is the tension of the unstable D-brane and $\be$ is a
constant. The potential has the minimum value $V(T=\pm\infty)=0$ and
the maximum value $V(T=0)=V_m$, with $V_m$ equal to the tension of
the unstable D$3$-brane. At the end of condensation,
lower-dimensional defects (stable D-branes and anti-D-branes) form
in regions where $T(\vec{x})=0$ and pressureless tachyon matter
forms in regions where $T(\vec{x})\rightarrow\pm\infty$ (the former
can also be viewed as the pressureless tachyon matter).

However, this tachyon cosmological model is examined to be
problematic \cite{Frolov:2002rr,Shiu:2002qe,Kofman:2002rh}. At early
times, the tachyon potential is too steep to produce large enough
e-foldings \cite{Frolov:2002rr,Shiu:2002qe,Kofman:2002rh}. As
tachyon rolls down the potential, we get a tachyon matter dominated
universe as the tachyon grows to infinity
\cite{Sen:2002in,Sen:2002an,Frolov:2002rr}. The pressureless tachyon
matter does not oscillate towards the minimum of the potential so
that ordinary tachyonic (p)reheating mechanism does not work. What
is more, we also encounter singular caustic formation at late times
when considering inhomogeneous condensation around a homogeneously
condensing tachyon field
\cite{Felder:2002sv,Felder:2004xu,Barnaby:2004nk,Hindmarsh:2009hs}.

In this paper, we try to improve the original tachyon cosmological
model by slightly modifying the tachyon potential. It is known that
the effective tachyon potential $V(T)$ should tend to zero as
$T\rightarrow\pm\infty$ when the closed string vacuum is reached,
according to Sen's conjecture (see review \cite{Sen:2004nf}). This
conjecture is now taken as a fact and has been verified in string
field theory
\cite{Sen:1999xm,deMelloKoch:2000ie,Moeller:2000jy,deMelloKoch:2000xf,
Moeller:2000hy,Berkovits:2000zj,Berkovits:2000zj,Berkovits:2000hf,
DeSmet:2000dp,Iqbal:2000st} in high accuracy (up to $90\%$ of the
conjectured value is achieved in superstring field theory).
Nevertheless, we consider the tachyon cosmology in the case when the
minimum of the potential is non-zero. A simple way to get this kind
of tachyon potential is to adding a positive constant $v$ into the
original potential $V$, so that the action for an unstable
three-brane becomes
\begin{equation}\label{e:CosMTDBIA}
 S=-\int d^{4}x\sqrt{-g}\left[V(T)+v\right]\sqrt{1+
g^{\mu\nu}\partial_\mu T\partial_\nu T}.
\end{equation}
When $T$ is small at the beginning, the potential $V(T)+v\sim V(T)$
(for small $v$). When $T$ grows large at late times, we have
$V(T)\sim0$ so that $V(T)+v\sim v$. Hence, the following features
are a natural result of this modified tachyon cosmological model:

\begin{itemize}

\item The effective potential is flattened, though the improvement is
limited for small $v$.

\item The final products of tachyon condensation in the new theory are
lower-dimensional branes (defects) within a stable 3-brane. By
contrast, in the original theory with $v=0$, the tachyon
condensation leads to defects and closed string vacuum.

\item We still have nearly tachyon matter for small $v$. But now it does
not generically tend to ``freeze" or develop caustics towards the
end of condensation, which are a generic result in all regions in
the original tachyon field theory.

\item The (p)reheating mechanisms adopted in the original theory still
work in the new theory. The fluctuations now do not vanish since the
descendent 3-brane can fluctuate.

\item After undergoing the inflationary, (p)reheating and
tachyon matter-dominated stages, the universe on the 3-brane ends
with a Chaplygin gas (CG) dark matter and dark energy stage (it does
not take infinitely long time to get this stage in the cosmological
background, though the minimum potential is at
$|T|\rightarrow\infty$). The constant $v$ contributes as the
cosmological constant in the end. Correspondingly, the ``velocity"
of the 3-brane grows first to a maximum value at a critical time and
then decreases, ultimately to zero.

\item All the above evolutionary processes of the universe happen on
the same 3-brane, which evolves from an unstable state to a stable
state. We do not need to compactify extra dimensions or annihilate
extra branes. In previous brane and tachyon inflation models, our
universe may reside on the lower-dimensional defects or on a
surviving brane from collisions of a stack brane-anti-brane.

\end{itemize}

Hence, the case with $v\neq0$ (even with very small $v$) is
completely different from the original theory with $v=0$: they lead
to different final products. The $v\neq0$ case leads to branes
within branes, though this kind of theory and potential has not yet
been constructed in string theory. This provides almost the simplest
scalar field model to date that can give a unified description of
the key features of the whole universe history.

The paper is organised as follows. We will examine the properties
and dynamics of the modified tachyon field theory
(\ref{e:CosMTDBIA}) in Minkowski spacetime in Section.\
\ref{sec:Min}. We further discuss the tachyon cosmology in this
modified theory in Section.\ \ref{sec:Tcos}. In Section.\
\ref{sec:Gen}, the generalised model that can give rise to the
generalised Chaplygin gas model is given. Conclusions are made in
the last section.

\section{Analysis in Minkowski spacetime}
\label{sec:Min}

The equation of motion from the action (\ref{e:CosMTDBIA}) is
\begin{equation}\label{e:EoM}
 \left[\Box T-f(T)\right](1+\pa T\cdot\pa T)
=\frac{1}{2}\partial^\mu T\partial_\mu(1+\pa T\cdot\pa T),
\end{equation}
where $\Box=g^{\mu\nu}\pa_\mu\pa_\nu$ and
\begin{equation}\label{e:}
 f(T)=\frac{V'}{V+v}=[\ln(V+v)]'.
\end{equation}
The primes denote derivative with respect to the field $T$. For the
potential (\ref{e:pot}), the expression is
\begin{equation}\label{e:}
 f(T)=-\frac{\be\tanh(\be T)}{1+\frac{v}{V_m}\cosh(\be
T)}.
\end{equation}
For $v=0$, $f(T)$ decreases monotonically from $f(T=0)=0$ to
$f(T\rightarrow\infty)=-\be$. For $v\neq0$, the situation is
different: $f$ first decreases from $f(T=0)=0$ to the minimum value
$f_{\textrm{min}}$ at some finite $T=T_{\textrm{m}}$, and then
increases to $f(T\rightarrow\infty)=0$. For a small $v$, we
approximately have
\begin{equation}\label{e:Tfmin}
 T_{\textrm{m}}\simeq\frac{1}{\be}\cosh^{-1}\left(\frac{V_m}{v}
\right)^{\frac{1}{3}}, \textrm{ }\textrm{ }\textrm{ }
f_{\textrm{min}}\simeq-\frac{\be}{1+\left(\frac{v}{V_m}
\right)^{\frac{2}{3}}}.
\end{equation}

In the original theory (\ref{e:TDBIA}) with $v=0$, there form
lower-dimensional defects in regions where $T=0$, which are kinks
and anti-kinks for the real tachyon field case, representing BPS
D2-brane and anti-D2-branes respectively. In the new theory with
$v\neq0$, kinks and anti-kinks should also form at $T=0$ because the
behaviour is similar near the top of the potential $V(T)$ and the
modified one $V(T)+v$.

However, in other regions where the tachyon rolls down the
potential: $V(T)\rightarrow0$, the action (\ref{e:CosMTDBIA}) will
evolve into the action describing a stable brane with tension $v$,
with the tachyon $T$ becoming a massless scalar $Y$:
\begin{equation}\label{e:CosYDBIA}
 S(T)\longrightarrow S(Y)=-v\int d^4x\sqrt{-g}
\sqrt{1+g^{\mu\nu}\partial_\mu Y\partial_\nu Y}.
\end{equation}
The scalar $Y$ describes the movement and fluctuations of the stable
D3-brane.

Hence, the final products of the condensation of the unstable
3-brane described by the theory (\ref{e:CosMTDBIA}) are a stable
3-brane plus lower-dimensional stable branes. The latter form inside
the 3-brane and may evolve further. We can understand this by taking
the unstable 3-brane described by (\ref{e:CosMTDBIA}) as a non-BPS
D3-brane with tension $V_m$ ``glued" to another BPS D3-brane with
tension $v$. The two branes share the same dynamics and are
described the unique action (\ref{e:CosMTDBIA}). As the tachyon
rolls down the potential, the ``non-BPS brane" will decay into lower
dimensional defects as usual, while the ``BPS brane" remain as the
final stable 3-brane.

In what follows, we examine the theory in detail in the Minkowski
spacetime, which indeed verifies the above speculations.

\subsection{Homogeneous solution}

In the homogeneous case, the energy density is conserved and is a
constant (no energy loss is assumed):
\begin{equation}\label{homT00}
 T_{00}=\frac{V+v}{\sqrt{1-\dot{T}^2}}.
\end{equation}
This case is similar to that in the original theory ($v=0$). We have
the minimum velocity $\dot{T}$ at $T=0$ and the maximum velocity
$|\dot{T}|=|\dot{T}|_{\textrm{max}}$ at $|T|\rightarrow\infty$. If
the minimum velocity is set to be $\dot{T}|_{T=0}=0$, it is easy to
determine
\begin{equation}\label{e:hommaxdotT}
 |\dot{T}|_{\textrm{max}}=\left[1-\left(\frac{v}{V_m+v}
\right)^2\right]^{\frac{1}{2}}.
\end{equation}
As expected, $|\dot{T}|_{\textrm{max}}=1$ when $v=0$. The brane is
driven by the tachyon potential to acquire the maximum velocity
$|\dot{T}|_{\textrm{max}}<1$ for $v\neq0$ at
$T\rightarrow\pm\infty$.

\subsection{Static solution}

In the static and 1-dimensional case (along the $x$-direction), the
pressure is conserved
\begin{equation}\label{}
 T_{11}=-\frac{V+v}{\sqrt{1+{T'}^2}}=-(V_0+v).
\end{equation}
where $T'=\pa_xT$ and $V_0=V(T_0)$. $T=T_0$ is where we get
$T'|_{T=T_0}=0$, i.e., where we get the maximum $|T|$. Thus, we have
the solution
\begin{equation}\label{}
 T'=\pm\left[\left(\frac{V+v}{V_0+v}
\right)^2-1\right]^{\frac{1}{2}}.
\end{equation}
In the $v=0$ case, this is the kink and anti-kink solution. The kink
and anti-kink are located at $T=0$. At the end of condensation
$V_0\rightarrow0$, the (anti-)kink solution becomes solitonic with
the field gradient $|T'|\rightarrow\infty$, representing a BPS
(anti-)D2-brane. For the $v\neq0$ case, we also have the maximum
field gradient at $T=0$ and
\begin{equation}\label{}
 |T'|_{\textrm{max}}=\left[\left(\frac{V_m+v}{v}
\right)^2-1\right]^{\frac{1}{2}} \textrm{ }\textrm{ as }\textrm{ }
V_0\rightarrow0.
\end{equation}
It looks like that the unstable 3-brane is rotated away from the
$x$-direction by an angle $\th=\tan^{-1}|T'|_{\textrm{max}}$ at the
``kink" position $T=0$ when it evolves into the stable 3-brane. In
the limit $v=0$, the angle is $\th=\pi/2$. This indicates that the
descendent BPS (anti-)D2-brane is a point-like object along the
$x-$dimension at the end of condensation in this case.

The energy of this system is
\begin{equation}\label{}
 \mathcal{E}=\int d^3x (V+v)\sqrt{1+{T'}^2}
=\int d^2x\int dx\frac{(V+v)^2}{V_0+v}.
\end{equation}
It is easy to prove that it is divergent for any $v\neq0$ in the
limit $V_0\rightarrow0$. It converges only when $v=0$, giving the
tension of the descendent D2-brane: $\pi V_m/\be$. That is, in the
$v\neq0$ case, we get an extended object, but not a solitonic
defect, along the $x$-direction.

\subsection{Spacetime-dependent condensation}

In the original tachyon effective theory (\ref{e:TDBIA}), both the
numerical and theoretical calculation in the spacetime-dependent
case indicate that there are singularities either near the core of
the defects \cite{Cline:2003vc,Barnaby:2004dz} or around extrema
(caustics)
\cite{Felder:2002sv,Felder:2004xu,Barnaby:2004nk,Hindmarsh:2009hs}.
It is suggested that they come from the fact that higher derivative
terms are not taken into account in the effective theory
\cite{Cline:2003vc}. We shall show that, in the new theory with the
modified potential, the former singularity still exists while the
latter is avoided.

As stated above, the dynamics near the top of the potential with
small $v\neq0$ is not altered dramatically compared with the $v=0$
case. In the vicinity of a position where $T=0$, we can expand the
field in the following way: $T(t,x)=a(t)x+(1/6)c(t)x^3+\cdots$
\cite{Cline:2003vc,Hindmarsh:2009hs}. Since
$f(T)|_{T\simeq0}\simeq-\be^2 T/[1+(v/V_m)]$, we have from the
equation of motion (\ref{e:EoM}):
\begin{equation}\label{e:}
 \ddot{a}=\frac{2a\dot{a}^2+c}{1+a^2}+\widetilde{\be} a,
\end{equation}
where $\widetilde{\be}=\be/(1+v/V_m)^{1/2}$. Neglecting $c$, we get
the solution:
\begin{equation}\label{}
 a(t)=\frac{a_0}{\cos(\widetilde{\be}t)},
\end{equation}
where $a_0$ is a constant. Thus, the behaviour of (anti-)kinks is
similar to that in the original theory with $v=0$: the gradient
grows to infinity in a finite time.

In other regions with no kinks and anti-kinks forming, the tachyon
can grow to infinity driven by the potential
\cite{Felder:2002sv,Felder:2004xu,Barnaby:2004nk}. Around extrema in
these regions, we split the field into a homogeneous field plus
small perturbations: $T(t,\vec{x})=T_0(t)+\tau(t,\vec{x})$
\cite{Hindmarsh:2009hs}. The equation of motion (\ref{e:EoM}) up to
the quadratic order thus becomes
\begin{eqnarray}\label{e:extremaEoM}
 [\ddot{T}_0+f(1-\dot{T}_0^2)]+[\ddot{\tau}-2f
\dot{T}_0\dot{\tau}-(1-\dot{T}_0^2)\nabla^2\tau]+
[(f+\ddot{T}_0)\nabla\tau\cdot\nabla{\tau}
\\ \nonumber
-f\dot{\tau}^2-2 \dot{T}_0(\nabla\tau\cdot\nabla
\dot{\tau}-\dot{\tau} \nabla^2\tau)]=0.
\end{eqnarray}

For $v=0$, $f(T)\rightarrow-\be$ as $|T|\rightarrow\infty$. The
leading terms in the above equation indicate that the velocity gets
the maximum value $|\dot{T}_0|_{\textrm{max}}\rightarrow1$, with
$\ddot{T}_0\rightarrow0$, as $|T_0|\rightarrow\infty$. The precise
solution leads to the relation:
\begin{equation}\label{}
 1+\pa T\cdot\pa
T\simeq-2\dot{\tau}+(\nabla\tau)^2\rightarrow0,
\end{equation}
which causes caustic formation: $\tau\rightarrow\infty$ in some
regions, and causes the perturbations to ``freeze":
$\tau\rightarrow0$ in some other regions. The caustic formation is a
signal of instability.

Now for the $v\neq0$ case, $f(T)\rightarrow0$ as
$|T|\rightarrow\infty$. So we get the maximum velocity
$|\dot{T}_0|_{\textrm{max}}<1$ at the end of condensation since
$\ddot{T}_0\rightarrow0$. If we take
$|\dot{T}_0|\simeq|\dot{T}_0|_{\textrm{max}}$ (a constant) at late
times, Eq.\ (\ref{e:extremaEoM}) becomes
\begin{equation}\label{}
 [\ddot{\tau}-(1-\dot{T}_0^2)\nabla^2\tau]-2\dot{T}_0
(\nabla\tau\cdot\nabla \dot{\tau}-\dot{\tau}\nabla^2\tau)=0.
\end{equation}
Thus, it has the plane wave solution of a massless scalar:
\begin{equation}\label{}
 \tau\sim e^{i\om t+i\vec{k}\cdot\vec{\widetilde{x}}},
\end{equation}
where $\vec{\widetilde{x}}=\vec{x}/\sqrt{1-T_0^2}$ and
$\om^2=\vec{k}^2$. Hence, the tachyon now does not generically
freeze and develop caustics at the end of condensation in the
modified theory. This is because we get a stable 3-brane at late
times. However, since a solution satisfying $1+\pa T\cdot\pa T=0$ is
always a solution to the full equation of motion (\ref{e:EoM}),
regardless of what the potential is, caustics may form at some
special positions where the spacetime-dependent solution happens to
satisfy $1+\pa Y\cdot\pa Y=0$. This is also realised in the
cosmological background in the Chaplygin gas model
\cite{Bilic:2001cg,Gorini:2007ta}.

So the preliminary analysis above is verified: the final products of
the tachyon condensation on this kind of unstable 3-branes are: a
stable 3-brane of tension $v$, containing lower-dimensional defects
(stable 2-brane pairs with opposite charges) or their descendent
objects.

\section{Tachyon cosmology with the modified potential}
\label{sec:Tcos}

We now consider the tachyon condensation process with the modified
potential in the FRW cosmological background. The energy density and
the pressure of this system are respectively:
\begin{eqnarray}\label{e:infenpre}
 \rho=\frac{V+v}{\sqrt{1-\dot{T}^2}}, &
 p=-(V+v)\sqrt{1-\dot{T}^2}.
\end{eqnarray}
The equation of state is
\begin{equation}\label{e:eos}
 w=\frac{p}{\rho}=-(1-\dot{T}^2).
\end{equation}
Thus, we can have pressureless tachyon matter only when
$|\dot{T}|=1$, which can occur in the $v=0$ case.

In a flat universe, the Friedmann equations are
\begin{equation}\label{e:friedman1}
 H^2=\frac{1}{3M_{Pl}^2}\frac{V+v}{\sqrt{1-\dot{T}^2}},
\end{equation}
and
\begin{equation}\label{e:friedman2}
 \frac{\ddot{a}}{a}=\frac{1}{3M_{Pl}^2}\frac{(V+v)(1-\frac{3}
{2}\dot{T}^2)}{\sqrt{1-\dot{T}^2}}.
\end{equation}
From the second equation, we learn that the accelerated expansion
occurs when $|\dot{T}|<\sqrt{6}/3$. If the maximum velocity is given
by Eq.\ (\ref{e:hommaxdotT}), the condition that we have a
decelerated stage after inflation is $v<V_m/(\sqrt{3}-1)$.

The equation for $T(t)$ in the expanding universe is
\begin{equation}\label{e:conservecond}
 \frac{\ddot{T}}{1-\dot{T}^2}+3H\dot{T}+f(T)=0,
\end{equation}
where $f=[\ln(V+v)]'$. So $\ln(V+v)$ is the ``effective" potential
that corresponds to the potential in ordinary scalar field model. It
is easy to verify that the addition of the constant $v\neq0$ tends
to flatten the original effective potential.

In the original tachyon condensation $v=0$, $f(T)$ decreases from
zero at $T=0$ to the minimum value $f_{\textrm{min}}(T)=-\be$ at
$T\rightarrow\infty$. In this situation, we get $\ddot{T}=0$ and the
maximum velocity $|\dot{T}|_{\textrm{max}}=1$ at the end of
condensation, as learned from Eq.\ (\ref{e:conservecond}). By
contrast, for the $v\neq0$ case, $f(T)$ reaches the minimum value
$f_{\textrm{min}}(T)$ at a finite field $T=T_{\textrm{min}}$ (as
given in Eq.\ (\ref{e:Tfmin}) for the small $v$ case) and then turns
to grow to zero. So it is possible that we get $3H\dot{T}+f(T)=0$ at
some critical value $T=T_c$ (different from $T_{\textrm{m}}$ in
Minkowski spacetime), where the velocity reaches the maximum value
$|\dot{T}|_{\textrm{max}}$. In this case, the velocity of the brane
constrained by Eq.\ (\ref{e:conservecond}) first grows to
$|\dot{T}|_{\textrm{max}}$ at $T=T_c$ and then decreases, ultimately
to zero as $T\rightarrow\infty$.

To illustrate this more precisely, we express Eq.\
(\ref{e:conservecond}) as follows, by inserting Eq.\
(\ref{e:friedman1})
\begin{equation}\label{e:conservecond2}
 \ddot{T}=-(1-\dot{T}^2)\sqrt{V+v}\left[\frac{\sqrt{3}}{M_{Pl}}
\frac{\dot{T}}{(1-\dot{T}^2)^{\frac{1}{4}}}+g(T)\right],
\end{equation}
where $g(T)=f(T)/\sqrt{V+v}\leq0$. Analogous to $f(T)$, the function
$g(T)$ equals to zero at $T=0,\infty$, and has a negative minimum
value at some value of $T$. This equation can be further simplified
to
\begin{equation}\label{e:conservecond3}
 \pa_t\left[\frac{(1-\dot{T}^2)^{\frac{1}{4}}}{\sqrt{V+v}}\right]
=\frac{\sqrt{3}}{2M_{Pl}}\dot{T}^2.
\end{equation}
The first term in the square bracket of Eq.\ (\ref{e:conservecond2})
increases monotonically with $\dot{T}$ as $T$ grows. At the
beginning, the term $g(T)$ dominates. As the tachyon rolls down the
potential, we shall get a critical point where the two terms inside
the square bracket cancels out, i.e.,
\begin{equation}\label{e:}
 T=T_c: \textrm{ }\textrm{ }\textrm{ } \ddot{T}=0.
\end{equation}
That is, we get the maximum velocity $|\dot{T}|_{\textrm{max}}$ at
this critical point. But this critical value $T_c$ takes a
complicated form generically and can not be easily determined
analytically. For the small $v$ case, $|\dot{T}|_{\textrm{max}}$ is
closed to 1 and so it is given by
\begin{equation}\label{e:}
 |\dot{T}|_{\textrm{max}}=\dot{T}|_{T=T_c}\simeq\left.\sqrt{1-
\frac{9(V+v)^6}{M_{Pl}^4{V'}^4}}\right|_{T=T_c}.
\end{equation}
Beyond this critical value $T_c$, the first term in the squared
bracket becomes dominant and the velocity $\dot{T}$ turns to
decrease, till to be zero, which can also been seen from Eq.
(\ref{e:conservecond3}).

Still another way to see this evolutionary feature of the velocity
is to analyze the action (\ref{e:CosMTDBIA}) itself. At the
beginning, $\dot{T}$ increases driven by the asymptotic potential
$V+v$. As $V(T)$ tends to zero at late times, the whole potential
$V+v$ becomes dominant by the constant $v$ and we get the action
(\ref{e:CosYDBIA}) for a stable 3-brane, with tahcyon $T$ evolving
into a massless scalar $Y$. It is known that the action
(\ref{e:CosYDBIA}) in the cosmological background leads to the
Chaplygin gas dark matter and dark energy universe
\cite{Kamenshchik:2001cp,Bilic:2001cg,Bilic:2002vm}. The expansion
of the Chaplygin gas universe is driven by the energy released from
the deceleration of the stable 3-brane in the bulk, i.e., from the
decreasing of the velocity $\dot{T}$ or $\dot{Y}$ \cite{Li:2010kz}.
Thus, in between the increasing and decreasing stages of the
velocity, there must be a critical point $T=T_c$, at which
$|\dot{T}|$ gets the maximum value $|\dot{T}|_{\textrm{max}}$.

In summary, the velocity of the 3-brane changes with the following
rule in the FRW metric as time grows
\begin{equation}\label{e:velevo}
 \dot{T}=0\longrightarrow|\dot{T}|=|\dot{Y}|=|\dot{T}|_{
\textrm{max}}\longrightarrow\dot{Y}=0.
\end{equation}
This evolutionary rule is different from that in the Minkowski
spacetime case, in which the brane will never really get the maximum
velocity until $T\rightarrow\pm\infty$.

In terms of Eqs.\ (\ref{e:eos}) and (\ref{e:friedman2}), the
evolutionary feature (\ref{e:velevo}) of the brane velocity implies
that the universe will undergo inflation, deceleration
(matter-dominated) and acceleration (dark-energy dominated) stages
as time grows if $v$ is small enough compared with $V_m$. The
evolutionary history of the brane universe is summarised as follows:
\begin{eqnarray}
 \{\dot{T}=0\} \textrm{ Inflation}\rightarrow\textrm{radiation
dominated}\rightarrow \textrm{tachyon matter } \nonumber \\
\textrm{}\{|\dot{T}|=|\dot{T}|_{\textrm{max}}\} \textrm{ CG
matter}\rightarrow\textrm{CG dark energy } \{\dot{T}=\dot{Y}=0\}.
\end{eqnarray}
If $v$ is large, we may not have the deceleration stage. Then the
brane universe always accelerates in the whole evolutionary history.
In what follows, we only consider the small $v$ case.

\subsection{Inflation}

Tachyon inflation occurs on or near the top of the potential. Under
the slow-roll conditions: $\dot{T}^2\ll 1$ and $\ddot{T}\ll
3H\dot{T}$, we have with the modified potential
\begin{equation}\label{e:infeq1}
 H^2\simeq\frac{1}{3M_{Pl}^2}(V+v),\textrm{ }\textrm{ }\textrm{ }
\dot{T}\simeq-\frac{[\ln(V+v)]'}{3H}.
\end{equation}
The number of e-foldings is given by
\begin{equation}
 N(t)\equiv\int_t^{t_e}H(t)dt\simeq-\int_T^{T_e}\frac{3H^2
(V+v)}{V'}dT.
\end{equation}
The slow-roll parameters are
\begin{equation}
 \epsilon\simeq\frac{M_{Pl}^2}{3}\frac{{V'}^2}{(V+v)^3},
\textrm{ }\textrm{ }\textrm{ }
\eta\simeq\frac{M_{Pl}^2}{3}\left[\frac{3{V'}^2}{(V+v)^3}
-\frac{2V''}{(V+v)^2}\right].
\end{equation}
Thus, the addition of a constant $v$ tends to flatten the effective
potential and to produce larger e-foldings with the slow-parameters
more strongly suppressed. Of course, the improvements are limited in
the small $v$ case. The energy scale difficulty in tachyon inflation
pointed out in \cite{Kofman:2002rh} is not released. Tachyon driven
inflation may be only part of the whole inflationary epoch, as
suggested in \cite{Kofman:2002rh,Sen:2003mv}.

Since there is very little change in the inflationary equations for
the small $v$ case, the discussion and predictions on inflation in
our model should be similar to the original tachyon inflation model
\cite{Gibbons:2002md,Fairbairn:2002yp,Choudhury:2002xu,Frolov:2002rr,
Kofman:2002rh,Piao:2002vf,Steer:2003yu}. Here, we do not copy the
results.

\subsection{Reheating and ``non-freezing" tachyon matter}

In the tachyon cosmological model, the universe is quickly dominated
by cold tachyon matter after inflation. Therefore, there should be
some mechanisms to reheat the universe and create particles. As
noted in \cite{Cline:2002it}, the tachyonic preheating mechanism
\cite{Felder:2000hj,Felder:2001kt} in hybrid inflation
\cite{Linde:1993cn} is not applicable in tachyon inflation with
$v=0$ because the tachyon does not oscillate as it rolls down the
potential. Therefore, other mechanisms via coupling to gauge field
\cite{Cline:2002it,Barnaby:2004dp} or by introducing the curvaton
field \cite{Lyth:2001nq,Campuzano:2005qw} are proposed to account
for the reheating and particle production in tachyon inflation.

In our case with $v\neq0$, the tachyon field evolves into a massless
scalar towards the end of condensation and the fluctuations do not
vanish, since the final product, the stable 3-brane, can fluctuate.
So this is more like the tachyonic preheating process in hybrid
inflation, compared with the $v=0$ case.

The reheating mechanism via coupling to gauge field proposed in
\cite{Cline:2002it,Barnaby:2004dp} also applies in our case with
small $v$. The DBI effective theory including gauge fields
\cite{Garousi:2000tr,Bergshoeff:2000dq,Kluson:2000iy}
\begin{equation}
 S=-\int d^4x(V+v)\sqrt{-\det(g_{\mu\nu}+\partial_\mu T
\partial_\nu T+F_{\mu\nu})}.
\end{equation}
As we have analyzed, there are also defects forming near $T=0$ in
the modified theory (\ref{e:CosMTDBIA}). The behaviour is similar to
the original case: $T=ax$, with $a$ tending to infinity at the end
of condensation. Around extrema, we also have the approximately
homogeneous field $T(t)$ plus small perturbations. When the part
$V(T)$ dominates the whole potential $V+v$, the discussion of
reheating follows that in the original theory made in
\cite{Cline:2002it,Barnaby:2004dp}. At late times when
$V\rightarrow0$ and the constant $v$ dominates, the gauge field will
decouple from the $T$ field in leading order. Meanwhile, the
universe enters into the Chaplygin gas dark universe as $v$ starts
to dominate the whole potential.

The velocity $\dot{T}$ reaches the maximum
$|\dot{T}|_{\textrm{max}}$ at a finite critical value $T=T_c$ (i.e.,
before $V=0$ is really reached). Approaching this critical value, we
get the ``tachyon matter" (not really pressureless) if $v$ is small.
As analyzed in the previous section, the tachyon matter does not
generically ``freeze" or develop cuastics
\cite{Felder:2004xu,Barnaby:2004nk,Hindmarsh:2009hs} because it will
become the Chaplygin gas dark matter beyond the critical value
$T=T_c$ (caustics may form in special case in the Chaplygin gas
model \cite{Bilic:2001cg,Gorini:2007ta}). Hence, we get rid (or
partially) of the instability problem \cite{Frolov:2002rr} and the
over-abundance problem \cite{Shiu:2002qe,Shiu:2002xp} about tachyon
matter found in the original theory: (i) the perturbations $\de
T=\tau$ of the tachyon turn to oscillating modes (since the
descendent stable 3-brane can fluctuate) instead of the growing mode
$\tau\propto t$ (which causes the instability) in the end; (ii) we
get less tachyon matter for a non-vanishing $v$ and even no tachyon
matter for a large enough $v$.

\subsection{Final stage: Chaplygin gas dark universe}

As $V$ tends towards $0$ and the constant $v$ dominates the
potential, the action (\ref{e:CosMTDBIA}) of an unstable 3-brane
evolves into the one (\ref{e:CosYDBIA}) for a stable 3-brane. We
enter into the Chaplygin gas-like universe
\cite{Kamenshchik:2001cp,Bilic:2001cg,Bilic:2002vm} with energy
density $\rho=v/\sqrt{1-\dot{T}^2}$ and pressure
$p=-v\sqrt{1-\dot{T}^2}$ given in Eq.\ (\ref{e:infenpre}),
satisfying:
\begin{equation}
 p=-\frac{v^2}{\rho}.
\end{equation}
We have discussed the evolutionary process in this stage in detail
in our previous work \cite{Li:2010kz}.

After the brane acquires this maximum velocity
$|\dot{T}|_{\textrm{max}}$, it turns to slow down due to some
mechanisms like gravitational waves linkage into the bulk, fueling
the expansion of the universe. The universe will undergo a matter
dominated stage and a second-time acceleration stage as the velocity
decreases. For very small $v$ (compared with $V_m$), the universe is
dominated by the Chaplygin gas dark matter with $\om\simeq0$ and
with nearly vanishing pressure. However, the perturbations of the
field in the Chaplygin gas universe can not account for CMB
observations \cite{Sandvik:2002jz,Amendola:2003bz} (for a review see
\cite{Copeland:2006wr}). Like the tachyon matter, the perturbations
of the Chaplygin gas dark matter behave as cold dark matter. But
they disappear in the acceleration stage.

When the velocity decreases further to be lower than
$|\dot{T}|=|\dot{Y}|<\sqrt{6}/3$, the expansion of the universe
turns to accelerate. As $\dot{Y}\rightarrow0$, the brane universe
evolves into a dS universe with the cosmological constant
\begin{equation}
 \Lambda=\frac{v}{M_{Pl}^2}.
\end{equation}
Thus, the tension $v$ of the final stable 3-brane is determined by
the cosmological constant which should be a very small value.

\section{Generalised model}
\label{sec:Gen}

In this section, we give the tachyon field action that can give rise
to the generalised Chaplygin gas (GCG) dark universe
\cite{Bilic:2001cg,Bento:2002ps}, which agrees better with
observations
\cite{Sandvik:2002jz,Bento:2002yx,Amendola:2003bz,Gorini:2007ta}. In
the generalised Chaplygin gas model, the energy density and pressure
are related via the relation: $p\rho^\alpha=-A$ $(0<\al<1)$.
Accordingly, the generalised tachyon field action can be taken as
\begin{equation}\label{e:genTaction}
 S=-\int d^4x\sqrt{-g}(V+v)\left(1-|\dot{T}|^
{\frac{1+\alpha}{\alpha}}\right)^{\frac{\alpha}{1+\alpha}}.
\end{equation}
Similar generalisation of the DBI action has been discussed in
\cite{Brax:2003rs}. In the homogeneous case, the maximum speed
corresponding to (\ref{e:hommaxdotT}) in the previous case is
\begin{equation}\label{e:genTmax}
 |\dot{T}|_{\textrm{max}}=\left[1-\left(\frac{v}{V_m+v}\right)^
{1+\al}\right]^{\frac{\alpha}{1+\alpha}}.
\end{equation}

In the FRW cosmology, the energy density and pressure of the system
are respectively
\begin{equation}\label{e:}
 \rho=(V+v)\left(1-|\dot{T}|^
{\frac{1+\alpha}{\alpha}}\right)^{-\frac{1}{1+\alpha}}, \textrm{
}\textrm{ }\textrm{ }
 p=-(V+v)\left(1-|\dot{T}|^
{\frac{1+\alpha}{\alpha}}\right)^{\frac{\alpha}{1+\alpha}}.
\end{equation}
The equation of state is
\begin{equation}
 \omega=-1+|\dot{T}|^{\frac{1+\alpha}{\alpha}}.
\end{equation}
Towards the end of condensation, they satisfy \footnote{If we want
to keep a regular form of the relation: $p\rho^\alpha=-v^2$, we can
express the potential in the action (\ref{e:genTaction}) as
$(V+v)^{2/(1+\alpha)}$. However, the potential profile in this form
is much sharper for $0<\al<1$ than in the $\al=1$ case. But such a
choice of the potential is good for the $\al>1$ case which is
suggested in \cite{Bertolami:2004ic} based on supernova data
analysis.}:
\begin{equation}
 p=-\frac{v^{1+\al}}{\rho^\alpha},
\end{equation}
which is just the relation between pressure and energy density in
the generalised Chaplygin gas model: $p\rho^\alpha=-A$ with
$A=v^{1+\al}$.

Thus, the Friedmann equations are
\begin{equation}\label{e:}
 H^2=\frac{V+v}{3M_{Pl}^2}\left(1-|\dot{T}|^
{\frac{1+\alpha}{\alpha}}\right)^{-\frac{1}{1+\alpha}}.
\end{equation}
\begin{equation}\label{e:}
 \frac{\ddot{a}}{a}=\frac{V+v}{3M_{Pl}^2}\left(1-|\dot{T}|^
\frac{\al+1}{\al}\right)^{-\frac{1}{\al+1}}\left(1-\frac{3}
{2}|\dot{T}|^\frac{\al+1}{\al}\right).
\end{equation}
If the maximum velocity is (\ref{e:genTmax}), then the condition
that we have a deceleration stage in between two acceleration stages
is: $v<V_m/(3^{1/(1+\al)}-1)$.

The equation of the scalar (for non-negative $\dot{T}$) can be
expressed as
\begin{equation}\label{e:}
 \left[\frac{\ddot{T}}{\al(1-\dot{T}^{\frac{1+\al}{\al}})}
+3H\dot{T}\right]\dot{T}^{\frac{1-\al}{\al}}+f(T)=0,
\end{equation}
where $f(T)$ is defined as before. It is easy to know that the
velocity $|\dot{T}|$ increases first and reaches the maximum value
at a finite critical value $T=T_c$. After this point, $|\dot{T}|$
turns to decrease to zero and we enter into the GCG dark universe.

Under the slow-roll conditions: $\dot{T}^{(1+\al)/\al}\ll 1$ and
$\ddot{T}\ll 3H\dot{T}$, we have the similar inflationary equations
to Eq.\ (\ref{e:infeq1}): $H^2\simeq(V+v)/(3M_{Pl}^2)$ and
$\dot{T}^{1/\al}=-[\ln(V+v)]'/(3H)$.

\section{Conclusions}
\label{sec:conclusions}

The original tachyon cosmological model is somehow improved by
adding a positive constant $v$ into the asymptotic tachyon
potential. In particular, some difficulties related to reheating and
tachyon matter can be (partially) cured. Moreover, the universe in
this modified theory evolves into a Chaplygin gas dark universe at
late times. Hence, in this simple single-scalar model, the universe
history including the inflationary, radiation-dominated, (dark)
matter-dominated and dark energy-dominated stages can be naturally
reproduced, at least at the qualitative level. Note that the tachyon
matter and Chaplygin gas matter in the model take similarity and can
be viewed as one phase. The constant $v$ is the only parameter in
this model.. It is related to the cosmological constant and controls
all the evolutionary processes of the brane universe. An unstable
3-brane described by the action (\ref{e:CosMTDBIA}) with an
appropriate $v$ behaves as our universe automatically in the FRW
metric. It is interesting to construct such kind of configuration in
string theory in future study.

\bibliographystyle{JHEP}
\bibliography{b}

\providecommand{\href}[2]{#2}\begingroup\raggedright\begin{thebibliography}{10}

\bibitem{Gibbons:2002md}
G.~W. Gibbons, {\it Cosmological evolution of the rolling tachyon},  {\em Phys.
  Lett.} {\bf B537} (2002) 1--4,
  [\href{http://xxx.lanl.gov/abs/hep-th/0204008}{{\tt hep-th/0204008}}].

\bibitem{Fairbairn:2002yp}
M.~Fairbairn and M.~H.~G. Tytgat, {\it {Inflation from a tachyon fluid?}},
  {\em Phys. Lett.} {\bf B546} (2002) 1--7,
  [\href{http://xxx.lanl.gov/abs/hep-th/0204070}{{\tt hep-th/0204070}}].

\bibitem{Choudhury:2002xu}
D.~Choudhury, D.~Ghoshal, D.~P. Jatkar, and S.~Panda, {\it {On the cosmological
  relevance of the tachyon}},  {\em Phys. Lett.} {\bf B544} (2002) 231--238,
  [\href{http://xxx.lanl.gov/abs/hep-th/0204204}{{\tt hep-th/0204204}}].

\bibitem{Frolov:2002rr}
A.~V. Frolov, L.~Kofman, and A.~A. Starobinsky, {\it {Prospects and problems of
  tachyon matter cosmology}},  {\em Phys. Lett.} {\bf B545} (2002) 8--16,
  [\href{http://xxx.lanl.gov/abs/hep-th/0204187}{{\tt hep-th/0204187}}].

\bibitem{Kofman:2002rh}
L.~Kofman and A.~Linde, {\it Problems with tachyon inflation},  {\em JHEP} {\bf
  07} (2002) 004, [\href{http://xxx.lanl.gov/abs/hep-th/0205121}{{\tt
  hep-th/0205121}}].

\bibitem{Piao:2002vf}
Y.-S. Piao, R.-G. Cai, X.-m. Zhang, and Y.-Z. Zhang, {\it {Assisted tachyonic
  inflation}},  {\em Phys. Rev.} {\bf D66} (2002) 121301,
  [\href{http://xxx.lanl.gov/abs/hep-ph/0207143}{{\tt hep-ph/0207143}}].

\bibitem{Steer:2003yu}
D.~A. Steer and F.~Vernizzi, {\it {Tachyon inflation: Tests and comparison with
  single scalar field inflation}},  {\em Phys. Rev.} {\bf D70} (2004) 043527,
  [\href{http://xxx.lanl.gov/abs/hep-th/0310139}{{\tt hep-th/0310139}}].

\bibitem{Cline:2002it}
J.~M. Cline, H.~Firouzjahi, and P.~Martineau, {\it {Reheating from tachyon
  condensation}},  {\em JHEP} {\bf 11} (2002) 041,
  [\href{http://xxx.lanl.gov/abs/hep-th/0207156}{{\tt hep-th/0207156}}].

\bibitem{Barnaby:2004dp}
N.~Barnaby and J.~M. Cline, {\it {Creating the universe from brane-antibrane
  annihilation}},  {\em Phys. Rev.} {\bf D70} (2004) 023506,
  [\href{http://xxx.lanl.gov/abs/hep-th/0403223}{{\tt hep-th/0403223}}].

\bibitem{Sen:2002in}
A.~Sen, {\it Tachyon matter},  {\em JHEP} {\bf 07} (2002) 065,
  [\href{http://xxx.lanl.gov/abs/hep-th/0203265}{{\tt hep-th/0203265}}].

\bibitem{Sen:2002an}
A.~Sen, {\it Field theory of tachyon matter},  {\em Mod. Phys. Lett.} {\bf A17}
  (2002) 1797--1804, [\href{http://xxx.lanl.gov/abs/hep-th/0204143}{{\tt
  hep-th/0204143}}].

\bibitem{Shiu:2002qe}
G.~Shiu and I.~Wasserman, {\it Cosmological constraints on tachyon matter},
  {\em Phys. Lett.} {\bf B541} (2002) 6--15,
  [\href{http://xxx.lanl.gov/abs/hep-th/0205003}{{\tt hep-th/0205003}}].

\bibitem{Garousi:2000tr}
M.~R. Garousi, {\it Tachyon couplings on non-bps d-branes and dirac-born-infeld
  action},  {\em Nucl. Phys.} {\bf B584} (2000) 284--299,
  [\href{http://xxx.lanl.gov/abs/hep-th/0003122}{{\tt hep-th/0003122}}].

\bibitem{Bergshoeff:2000dq}
E.~A. Bergshoeff, M.~de~Roo, T.~C. de~Wit, E.~Eyras, and S.~Panda, {\it
  T-duality and actions for non-bps d-branes},  {\em JHEP} {\bf 05} (2000) 009,
  [\href{http://xxx.lanl.gov/abs/hep-th/0003221}{{\tt hep-th/0003221}}].

\bibitem{Kutasov:2003er}
D.~Kutasov and V.~Niarchos, {\it Tachyon effective actions in open string
  theory},  {\em Nucl. Phys.} {\bf B666} (2003) 56--70,
  [\href{http://xxx.lanl.gov/abs/hep-th/0304045}{{\tt hep-th/0304045}}].

\bibitem{Felder:2002sv}
G.~N. Felder, L.~Kofman, and A.~Starobinsky, {\it Caustics in tachyon matter
  and other born-infeld scalars},  {\em JHEP} {\bf 09} (2002) 026,
  [\href{http://xxx.lanl.gov/abs/hep-th/0208019}{{\tt hep-th/0208019}}].

\bibitem{Felder:2004xu}
G.~N. Felder and L.~Kofman, {\it Inhomogeneous fragmentation of the rolling
  tachyon},  {\em Phys. Rev.} {\bf D70} (2004) 046004,
  [\href{http://xxx.lanl.gov/abs/hep-th/0403073}{{\tt hep-th/0403073}}].

\bibitem{Barnaby:2004nk}
N.~Barnaby, {\it {Caustic formation in tachyon effective field theories}},
  {\em JHEP} {\bf 07} (2004) 025,
  [\href{http://xxx.lanl.gov/abs/hep-th/0406120}{{\tt hep-th/0406120}}].

\bibitem{Hindmarsh:2009hs}
M.~Hindmarsh and H.~Li, {\it {Inhomogeneous tachyon condensation}},  {\em JHEP}
  {\bf 06} (2009) 050, [\href{http://xxx.lanl.gov/abs/0903.2019}{{\tt
  arXiv:0903.2019}}].

\bibitem{Sen:2004nf}
A.~Sen, {\it Tachyon dynamics in open string theory},  {\em Int. J. Mod. Phys.}
  {\bf A20} (2005) 5513--5656,
  [\href{http://xxx.lanl.gov/abs/hep-th/0410103}{{\tt hep-th/0410103}}].

\bibitem{Sen:1999xm}
A.~Sen, {\it {Universality of the tachyon potential}},  {\em JHEP} {\bf 12}
  (1999) 027, [\href{http://xxx.lanl.gov/abs/hep-th/9911116}{{\tt
  hep-th/9911116}}].

\bibitem{deMelloKoch:2000ie}
R.~de~Mello~Koch, A.~Jevicki, M.~Mihailescu, and R.~Tatar, {\it {Lumps and
  p-branes in open string field theory}},  {\em Phys. Lett.} {\bf B482} (2000)
  249--254, [\href{http://xxx.lanl.gov/abs/hep-th/0003031}{{\tt
  hep-th/0003031}}].

\bibitem{Moeller:2000jy}
N.~Moeller, A.~Sen, and B.~Zwiebach, {\it {D-branes as tachyon lumps in string
  field theory}},  {\em JHEP} {\bf 08} (2000) 039,
  [\href{http://xxx.lanl.gov/abs/hep-th/0005036}{{\tt hep-th/0005036}}].

\bibitem{deMelloKoch:2000xf}
R.~de~Mello~Koch and J.~P. Rodrigues, {\it {Lumps in level truncated open
  string field theory}},  {\em Phys. Lett.} {\bf B495} (2000) 237--244,
  [\href{http://xxx.lanl.gov/abs/hep-th/0008053}{{\tt hep-th/0008053}}].

\bibitem{Moeller:2000hy}
N.~Moeller, {\it {Codimension two lump solutions in string field theory and
  tachyonic theories}},  \href{http://xxx.lanl.gov/abs/hep-th/0008101}{{\tt
  hep-th/0008101}}.

\bibitem{Berkovits:2000zj}
N.~Berkovits, {\it {The tachyon potential in open Neveu-Schwarz string field
  theory}},  {\em JHEP} {\bf 04} (2000) 022,
  [\href{http://xxx.lanl.gov/abs/hep-th/0001084}{{\tt hep-th/0001084}}].

\bibitem{Berkovits:2000hf}
N.~Berkovits, A.~Sen, and B.~Zwiebach, {\it {Tachyon condensation in
  superstring field theory}},  {\em Nucl. Phys.} {\bf B587} (2000) 147--178,
  [\href{http://xxx.lanl.gov/abs/hep-th/0002211}{{\tt hep-th/0002211}}].

\bibitem{DeSmet:2000dp}
P.-J. De~Smet and J.~Raeymaekers, {\it {Level four approximation to the tachyon
  potential in superstring field theory}},  {\em JHEP} {\bf 05} (2000) 051,
  [\href{http://xxx.lanl.gov/abs/hep-th/0003220}{{\tt hep-th/0003220}}].

\bibitem{Iqbal:2000st}
A.~Iqbal and A.~Naqvi, {\it {Tachyon condensation on a non-BPS D-brane}},
  \href{http://xxx.lanl.gov/abs/hep-th/0004015}{{\tt hep-th/0004015}}.

\bibitem{Cline:2003vc}
J.~M. Cline and H.~Firouzjahi, {\it Real-time d-brane condensation},  {\em
  Phys. Lett.} {\bf B564} (2003) 255--260,
  [\href{http://xxx.lanl.gov/abs/hep-th/0301101}{{\tt hep-th/0301101}}].

\bibitem{Barnaby:2004dz}
N.~Barnaby, A.~Berndsen, J.~M. Cline, and H.~Stoica, {\it Overproduction of
  cosmic superstrings},  {\em JHEP} {\bf 06} (2005) 075,
  [\href{http://xxx.lanl.gov/abs/hep-th/0412095}{{\tt hep-th/0412095}}].

\bibitem{Kamenshchik:2001cp}
A.~Y. Kamenshchik, U.~Moschella, and V.~Pasquier, {\it {An alternative to
  quintessence}},  {\em Phys. Lett.} {\bf B511} (2001) 265--268,
  [\href{http://xxx.lanl.gov/abs/gr-qc/0103004}{{\tt gr-qc/0103004}}].

\bibitem{Bilic:2001cg}
N.~Bilic, G.~B. Tupper, and R.~D. Viollier, {\it {Unification of dark matter
  and dark energy: The inhomogeneous Chaplygin gas}},  {\em Phys. Lett.} {\bf
  B535} (2002) 17--21, [\href{http://xxx.lanl.gov/abs/astro-ph/0111325}{{\tt
  astro-ph/0111325}}].

\bibitem{Bilic:2002vm}
N.~Bilic, G.~B. Tupper, and R.~D. Viollier, {\it {Dark matter, dark energy and
  the Chaplygin gas}},  \href{http://xxx.lanl.gov/abs/astro-ph/0207423}{{\tt
  astro-ph/0207423}}.

\bibitem{Li:2010kz}
H.~Li, {\it {Cosmological evolution of a D-brane}},  {\em Phys. Rev.} {\bf D83}
  (2011) 066002, [\href{http://xxx.lanl.gov/abs/1010.4099}{{\tt
  arXiv:1010.4099}}].

\bibitem{Sen:2003mv}
A.~Sen, {\it {Remarks on tachyon driven cosmology}},  {\em Phys. Scripta} {\bf
  T117} (2005) 70--75, [\href{http://xxx.lanl.gov/abs/hep-th/0312153}{{\tt
  hep-th/0312153}}].

\bibitem{Felder:2000hj}
G.~N. Felder {\em et.~al.}, {\it {Dynamics of symmetry breaking and tachyonic
  preheating}},  {\em Phys. Rev. Lett.} {\bf 87} (2001) 011601,
  [\href{http://xxx.lanl.gov/abs/hep-ph/0012142}{{\tt hep-ph/0012142}}].

\bibitem{Felder:2001kt}
G.~N. Felder, L.~Kofman, and A.~D. Linde, {\it {Tachyonic instability and
  dynamics of spontaneous symmetry breaking}},  {\em Phys. Rev.} {\bf D64}
  (2001) 123517, [\href{http://xxx.lanl.gov/abs/hep-th/0106179}{{\tt
  hep-th/0106179}}].

\bibitem{Linde:1993cn}
A.~D. Linde, {\it {Hybrid inflation}},  {\em Phys. Rev.} {\bf D49} (1994)
  748--754, [\href{http://xxx.lanl.gov/abs/astro-ph/9307002}{{\tt
  astro-ph/9307002}}].

\bibitem{Lyth:2001nq}
D.~H. Lyth and D.~Wands, {\it {Generating the curvature perturbation without an
  inflaton}},  {\em Phys. Lett.} {\bf B524} (2002) 5--14,
  [\href{http://xxx.lanl.gov/abs/hep-ph/0110002}{{\tt hep-ph/0110002}}].

\bibitem{Campuzano:2005qw}
C.~Campuzano, S.~del Campo, and R.~Herrera, {\it {Curvaton reheating in
  tachyonic inflationary models}},  {\em Phys. Lett.} {\bf B633} (2006)
  149--154, [\href{http://xxx.lanl.gov/abs/gr-qc/0511128}{{\tt
  gr-qc/0511128}}].

\bibitem{Kluson:2000iy}
J.~Kluson, {\it {Proposal for non-BPS D-brane action}},  {\em Phys. Rev.} {\bf
  D62} (2000) 126003, [\href{http://xxx.lanl.gov/abs/hep-th/0004106}{{\tt
  hep-th/0004106}}].

\bibitem{Shiu:2002xp}
G.~Shiu, S.~H.~H. Tye, and I.~Wasserman, {\it {Rolling tachyon in brane world
  cosmology from superstring field theory}},  {\em Phys. Rev.} {\bf D67} (2003)
  083517, [\href{http://xxx.lanl.gov/abs/hep-th/0207119}{{\tt
  hep-th/0207119}}].

\bibitem{Sandvik:2002jz}
H.~Sandvik, M.~Tegmark, M.~Zaldarriaga, and I.~Waga, {\it {The end of unified
  dark matter?}},  {\em Phys. Rev.} {\bf D69} (2004) 123524,
  [\href{http://xxx.lanl.gov/abs/astro-ph/0212114}{{\tt astro-ph/0212114}}].

\bibitem{Amendola:2003bz}
L.~Amendola, F.~Finelli, C.~Burigana, and D.~Carturan, {\it {WMAP and the
  Generalized Chaplygin Gas}},  {\em JCAP} {\bf 0307} (2003) 005,
  [\href{http://xxx.lanl.gov/abs/astro-ph/0304325}{{\tt astro-ph/0304325}}].

\bibitem{Copeland:2006wr}
E.~J. Copeland, M.~Sami, and S.~Tsujikawa, {\it {Dynamics of dark energy}},
  {\em Int. J. Mod. Phys.} {\bf D15} (2006) 1753--1936,
  [\href{http://xxx.lanl.gov/abs/hep-th/0603057}{{\tt hep-th/0603057}}].

\bibitem{Bento:2002ps}
M.~C. Bento, O.~Bertolami, and A.~A. Sen, {\it {Generalized Chaplygin gas,
  accelerated expansion and dark energy-matter unification}},  {\em Phys. Rev.}
  {\bf D66} (2002) 043507, [\href{http://xxx.lanl.gov/abs/gr-qc/0202064}{{\tt
  gr-qc/0202064}}].

\bibitem{Bento:2002yx}
M.~d.~C. Bento, O.~Bertolami, and A.~A. Sen, {\it {Generalized Chaplygin gas
  and CMBR constraints}},  {\em Phys. Rev.} {\bf D67} (2003) 063003,
  [\href{http://xxx.lanl.gov/abs/astro-ph/0210468}{{\tt astro-ph/0210468}}].

\bibitem{Gorini:2007ta}
V.~Gorini, A.~Y. Kamenshchik, U.~Moschella, O.~F. Piattella, and A.~A.
  Starobinsky, {\it {Gauge-invariant analysis of perturbations in Chaplygin gas
  unified models of dark matter and dark energy}},  {\em JCAP} {\bf 0802}
  (2008) 016, [\href{http://xxx.lanl.gov/abs/0711.4242}{{\tt
  arXiv:0711.4242}}].

\bibitem{Brax:2003rs}
P.~Brax, J.~Mourad, and D.~A. Steer, {\it Tachyon kinks on non bps d-branes},
  {\em Phys. Lett.} {\bf B575} (2003) 115--125,
  [\href{http://xxx.lanl.gov/abs/hep-th/0304197}{{\tt hep-th/0304197}}].

\bibitem{Bertolami:2004ic}
O.~Bertolami, A.~A. Sen, S.~Sen, and P.~Silva, {\it {Latest supernova data in
  the framework of Generalized Chaplygin Gas model}},  {\em
  Mon.Not.Roy.Astron.Soc.} {\bf 353} (2004) 329,
  [\href{http://xxx.lanl.gov/abs/astro-ph/0402387}{{\tt astro-ph/0402387}}].

\end{thebibliography}\endgroup

\end{document}